\documentclass[usegraphicx,useAMS,usenatbib]{mn2e}

\newcommand{\Mdotd}{\dot M_d}

\newcommand{\Mdotstar}{\dot M_*}

\newcommand{\Mdisk}{M_{\rm disc}}
\newcommand{\Msun}{M_{\odot}}

\newcommand{\Rhole}{R_{\rm hole}}
\newcommand{\Rinner}{R_{\rm inner}}
\newcommand{\Msunperyr}{M_{\odot}\,{\rm yr}^{-1}}
\newcommand{\Lsun}{L_{\odot}}

\newcommand{\kms}{{\rm km}\,{\rm s}^{-1}}

\def\micron{\hbox{$\mu$m}}

\title[Demographics of Transition Objects]{Demographics of Transition Objects}
\author[J. R. Najita, S. E. Strom, and J. Muzerolle]{Joan R. Najita$^{1}$, Stephen E. Strom$^{1}$ and James Muzerolle$^{2}$ \\ 
$^{1}$National Optical Astronomy Observatory, 950 N. Cherry Ave.,
Tucson, AZ 85719\\
$^{2}$Steward Observatory, 933 N. Cherry Ave., Tucson, AZ, 85721
}

\begin{document}

\date{Submitted 2006 November.}

\pagerange{\pageref{firstpage}--\pageref{lastpage}} \pubyear{2002}

\maketitle

\label{firstpage}

\begin{abstract}
The unusual properties of transition objects (young stars with
an optically thin inner disc surrounded by an optically thick outer
disc) suggest that significant disc evolution has occured in 
these systems.
We explore the nature of these systems by examining their demographics,
specifically their stellar accretion rates $\Mdotstar$ and
disc masses $\Mdisk$ compared to
those of accreting T Tauri stars of comparable age.
We find that transition objects in Taurus occupy a restricted region
of the $\Mdotstar$ vs.\ $\Mdisk$ plane.  Compared to non-transition
single stars in Taurus, they have stellar accretion rates that are
typically $\sim 10$ times lower at the same disc mass and
median disc masses $\sim 4$ times larger.
These properties are anticipated by several proposed 
planet formation theories and 
suggest that the formation of Jovian mass planets may play a significant 
role in explaining the origin of at least some transition objects. 
Considering transition objects as a distinct demographic group
among accreting T Tauri stars 
leads to a tighter relationship between
disc masses and stellar accretion rates, with a slope
between the two quantities that is close to the value of unity
expected in simple theories of disc accretion.
\end{abstract}


\section{Introduction}

Transition objects are an interesting class of young stellar objects
whose properties indicate that significant disc evolution has
occurred.
Their spectral energy distributions (SEDs) imply the presence of
an optically thin inner region (defined by an ``opacity hole'' of
radius $\Rhole \sim 1-20$\,AU) that is surrounded by an optically
thick outer disc (beyond $\Rhole$).
Such an SED indicates that the circumstellar disc has evolved
significantly from the radially continuous optically thick disc
that is believed to surround all stars at birth.

Transition objects were first identified from the analysis of
the SEDs of large samples of solar-like pre-main sequence
stars in nearby star-forming regions (e.g. Strom et al.\ 1989;
Skrutskie et al.\ 1990; Gauvin and Strom 1992; Marsh \& Mahoney 1992;
Wolk \& Walter 1996).
They were selected to have SEDs whose long wavelength excesses
($\lambda \ga 10\micron$) were equal to or exceeded that
expected from a
geometrically flat optically thick disc, and whose shorter
wavelength excesses were below those of an optically
thick disc that extends to within a few stellar radii.
Thus the definition included both objects with {\it no
dust} within some disc radius as well as objects with
{\it optically thin} inner dust discs.

Transitional SEDs are expected to arise via multiple pathways.
For example, a transitional SED could arise as grains in
the inner disc grow into larger bodies, thereby
decreasing the continuum optical depth of the inner disc
(Strom et al.\ 1989; Dullemond \& Dominik 2005).
Alternatively, as suggested by Skrutskie et al.\ (1990),
a transitional SED may arise through the creation of a
large gap, the result of the dynamical isolation
of the inner and outer discs by a sufficiently massive giant planet
(see also Marsh \& Mahoney 1992;
Calvet et al.\ 2002; Rice et al.\ 2003; D'Alessio et al.\ 2005;
Calvet et al.\ 2005; Quillen et al.\ 2004).
More recently, disc photoevaporation has been suggested as an
explanation for the properties of some transition objects
(Clarke, Gendrin \& Sotomayor 2001; 
Alexander, Clarke \& Pringle 2006; McCabe et al.\ 2006).
Because all of the above scenarios can in principle produce
similar SEDs, additional diagnostic tools beyond the SED may
be needed to determine the physical state of a transition
object.

To take steps in this direction, we therefore examine in this
contribution the demographics of transition objects.  Specifically,
we compare the stellar accretion rates and disc masses
of transition objects with those of accreting T Tauri stars of
comparable age.
Our methods and results are described in \S 2 and \S 3 respectively.
The implications of the results are discussed in \S 4 and
summarized in \S 5.

\section{Methods}

\subsection{Utility of a Demographic Approach}
In order to explore the nature of transition objects, we
take a demographic approach, examining the properties of
an ensemble of transition objects compared to the properties of
a parent population of classical T Tauri stars.  Specifically, we
compare two well-studied properties of these systems, their stellar
accretion rates and their disc masses with those of accreting
T Tauri stars of comparable age.
A demographic approach has the advantage of being robust against 
{\it systematic\/}
errors, since systematic errors (e.g., of the kind described by
White \& Hillenbrand 2004 and Hartmann et al.\ 2006)
can shift demographic trends up or down in $\Mdotstar$ or
$\Mdisk$ but should not obscure them.
Since the use of a large sample of T Tauri stars and transition objects
also allows us to ``average over'' random measurement uncertainties
associated with the individual measurements,
such an approach may prove useful in identifying demographic trends
and, thereby, provide clues to the physical state of transition objects.
As a result of significant
recent efforts to measure stellar accretion rates and disc masses
for young stars, measurements of both quantities are now available
for a relatively large number of sources in the Taurus star forming region.

\subsection{Stellar Accretion Rates}

We adopted stellar accretion rates from the literature, all of which
were determined using some measure of the UV/optical accretion
excess emission.  Stellar accretion rates thus derived are fairly
robust.  They have been used, for example, to demonstrate 
correlations between $\Mdotstar$ and stellar mass
(e.g., Muzerolle et al.\ 2003; Natta et al.\ 2004) as 
well as $\Mdotstar$ and age 
(e.g., Sicilia-Aguilar et al.\ 2005).

Stellar accretion rates (Table 1) were adopted from the following references,
in order of preference: Gullbring et al.\ (1998, hereafter, G98);
Hartmann et al.\ (1998, hereafter, H98); White \& Ghez (2001, hereafter, WG01);
Calvet et al.\ (2004); Gullbring et al.\ (2000);
Hartigan, Edwards \& Ghandour (1995, hereafter, HEG95).  Values from WG01 were used
for all binaries when available, since their observations were specifically
tailored for resolving most binary systems. 
Systematic differences between some of these determinations arise from several
factors, including different extinction estimates, bolometric corrections,
and pre-main sequence tracks adopted to estimate stellar masses.
We have thus attempted to place all the values on a consistent scale
by calculating any systematic offsets for objects with measurements
in common with H98 (which includes all values from G98).

We show two such comparisons with measurements from WG01
and HEG95 in Figure~1. One can see in
both cases a similar trend with a relatively small dispersion
of $\sim 0.2$ dex, which is, perhaps surprisingly, 
smaller than the typical 
$\sim 0.5$ dex measurement uncertainties discussed by G98
and the typical veiling variability of a factor of 2 or less
(e.g., Hartigan et al.\ 1991).  However, there are significant
systematic offsets.  The WG01 values are lower on average by
a factor of 2.  Since they use the same methodology as H98,
most of the offset is likely a result of their adoption of
a different set of pre-main sequence tracks (Baraffe et al.\ 1998; 
Palla \& Stahler 1999)
which give systematically higher stellar masses than the D'Antona \&
Mazzitelli (1998) tracks used by H98.  Meanwhile,
the HEG95 values are higher on average by a factor of 8.
This discrepancy, discussed in detail in G98, results from
differences in extinction estimates, bolometric corrections, 
and the assumed accretion geometry.  Based on these comparisons,
we have scaled the values from WG01 up by a factor of 2
and the values from HEG95 down by a factor of 0.12 to place them
on the same scale as H98.

Two objects, CoKu Tau/4 and UX Tau A, lack published UV measurements.
We have thus estimated their stellar accretion rates using magnetospheric
accretion models of H$\alpha$ profiles 
(e.g., Muzerolle, Calvet \& Hartmann 2001
and references therein).  The value for CoKu Tau/4 is an upper limit
based on the line equivalent width since no resolved profiles exist
in the literature.  
Based on the scatter in the Figure 1, we estimate that the 
resulting (random) uncertainty in the stellar accretion rates adopted
here is likely to be less than a factor of $\sim 3$ or so, much less
than the spread of $\sim 100$ in the stellar
accretion rate at a given disc mass that is 
found in Taurus (\S 3). 

\setcounter{figure}{0}
\begin{figure}
\includegraphics[width=84mm]{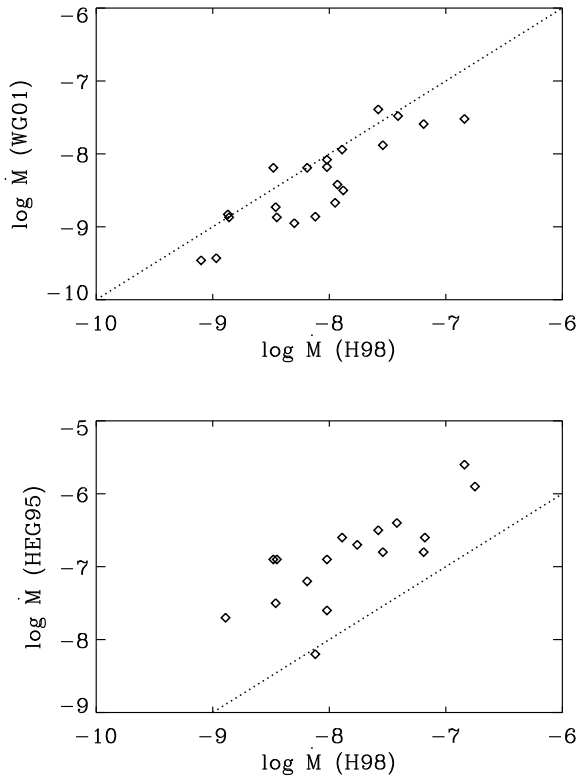}
\caption{Comparison of stellar accretion rates (in $\Msunperyr$) 
from Hartmann et al.\ (1998)
with those from White \& Ghez (2001) (top panel, including only single stars)
and Hartigan et al.\ (1995) (bottom panel).  Note the systematic offsets
in both cases, while the general trend with relatively small dispersion
is maintained.
}
\end{figure}

\subsection{Disk Masses}

We adopted the disc masses (Table 1) estimated for Taurus sources
based on submillimeter continuum measurements
(Andrews \& Williams 2005).
The data set has the advantage of being large and homogeneous.
While Andrews \& Williams (2005) estimated disc masses from SED fits where
possible, the disc masses for most of the sample were determined from an
empirical conversion between submillimeter fluxes and disc mass
based on the SED fits.  The random error in the disc masses is
expected to be $\la 0.5$\,dex, primarily the result of the
uncertainty in the spectral index at submillimeter wavelengths.
Uncertainties in this range are much less than the two orders of
magnitude spread in disc mass at a given stellar accretion rate
that is found in Taurus (\S 3).

Systematic errors are also possible.  For example,
Hartmann et al.\ (2006) suggest that disc masses are
likely to have been systematically underestimated because
maximum grain sizes 3 -- 10 times larger or smaller than 1 mm result in
grain opacities significantly lower than is typically assumed.
We can estimate the extent to which disc masses may be
underestimated by looking at the upper end of the disc mass
distribution reported by Andrews \& Williams.
Correction factors larger than 0.5 dex cannot be tolerated at
the upper mass end without driving the disc masses over the
theoretical gravitational instability limit. 
In any event, a demographic approach is robust against systematic errors 
in disk masses arising from under- or over-estimates of grain opacity,  
since systematic errors will shift demographic trends up or down in
$\Mdisk$ but will not obscure them.

\subsection{SED Classification}

In Table 1, we provide a rudimentary classification of
the observed SEDs of the Taurus sources based on
ground-based photometry and mid-IR spectra from Spitzer/IRS
(Furlan et al.\ 2006).  Objects were classified
as transition objects (T) if they show weak or no infrared excess
shortward of $10\,\mu$m and a significant excess at longer wavelengths.
The initial selection and classification was made by visual
inspection of the SED fits of Furlan et al.\ (2006).  This resulted
in our selecting objects whose excess emission fell by $>0.2$ dex 
below that of the ``Taurus Median'' obtained from the analysis 
of all of the class II sources examined by D'Alessio et al.\ (1999).
Systems with no significant infrared excess out to 
$\sim 30\,\mu$m, indicating no substantial disc material, were classified
as ``weak'' objects (W).  All other systems exhibit strong
excess emission indicative
of optically thick discs extending all the way in to the dust sublimation
radius, either ``classical'' discs (C) or ``flat-spectrum'' (F) sources with
elevated accretion rates and possible remnant envelopes.

Note that in several cases, the transition 
object classification may be influenced by the presence of unresolved
companions with possible infrared excesses of their own.  Also, for stars
with lower masses, even ``classical'' discs can often exhibit lower
excess emission at shorter wavelengths because the stellar irradiation flux,
which is typically the dominant source of dust heating, is generally smaller.
The likelihood of incorrectly classifying an object as 
``transition'' as opposed to ``classical'' is thus greater among
low mass, later spectral type stars owing to the decreased contrast
between disc and photosphere.   
We therefore limit our analysis to stars with spectral types earlier 
than M3. 
We have indicated the remaining ambiguous cases in Table 1 (C?, T?, C/T).
Table 2 captures the decrements from the Taurus median for all 
sources that have a transition-like SED (i.e., sources
with a T, T?, or C/T SED classification in Table 1).

Our classification scheme is broader than that adopted in some recent
studies in which only systems with {\it no} apparent excess anywhere
shortward of a given wavelength are considered to be transition objects
(Muzerolle et al.\ 2006; Sicilia-Aguilar et al.\ 2006).  
We accept a broader range of excesses as transition objects, including
those with weak excesses shortward of $10\micron$, in order to
avoid excluding systems in possibly interesting phases of planet formation.
For example,
systems that are dissipating their discs via photoevaporation
are expected to have essentially
no dust within $\Rhole$ and hence no excess above photospheric
levels at short wavelengths.  In contrast,
a significant continuum excess might be expected from discs that
have formed a Jovian mass planet and are replenishing,
via accretion streams, an inner disc with gas and {\it dust} from an outer disc.
In order to capture objects that are possibly in these states of evolution,
we selected objects that show a significant decrement ($\ge 0.2$\,dex)
relative to the Taurus median SED.

\setcounter{figure}{1}
\begin{figure}
\includegraphics[width=84mm]{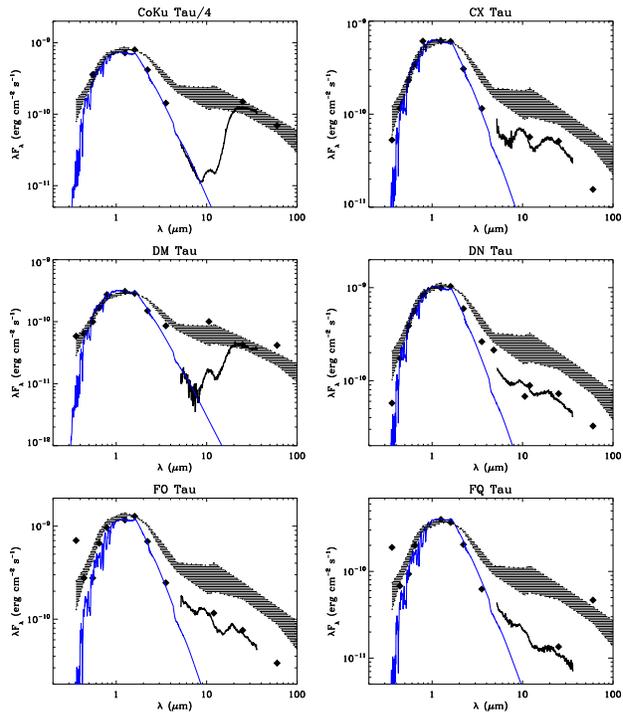}
\caption{SEDs for all stars earlier than M3 that we identify as transition 
objects.  Diamonds: optical and IRAS photometry from Kenyon \& Hartmann 
(1995); near-infrared photometry from 2MASS; IRAC photometry from 
Hartmann et al. (2005).  Black solid lines: IRS spectra from Furlan et al. (2006).  Blue solid lines: Kurucz model photospheres of the appropriate spectral type,
normalized to the dereddened J-band flux.  Hashed region: the upper and lower 
quartiles of the Taurus CTTS median SED from D'Alessio et al. (1999).
}
\end{figure}

\setcounter{figure}{1}
\begin{figure}
\includegraphics[width=84mm]{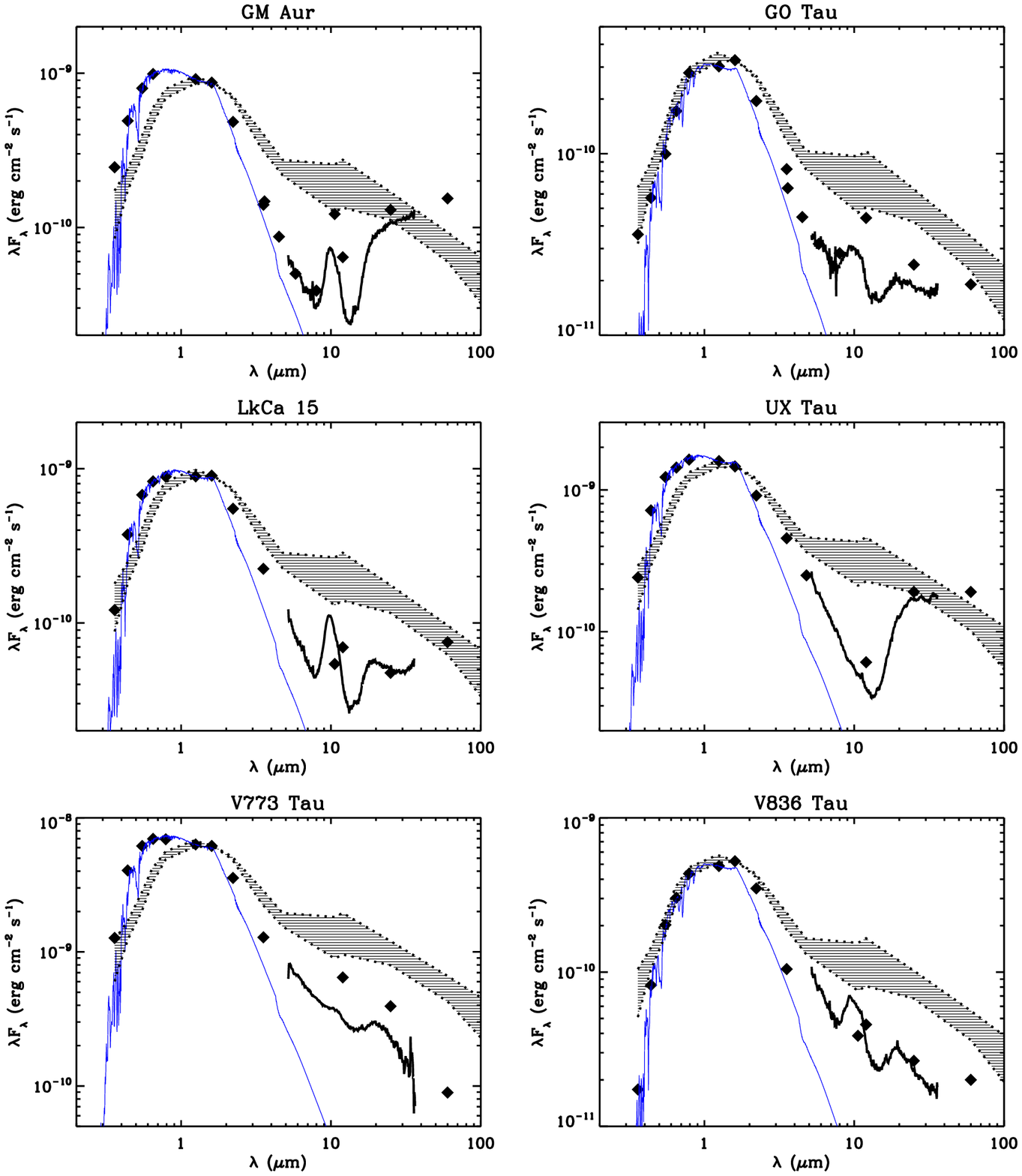}
\caption{{\it continued} Additional SEDs for objects identified as transition 
objects.}
\end{figure}

\section{Results}

Figure 3 plots the recalibrated stellar accretion rates from \S 2.2 against
the disc masses described in \S 2.3.
The top panel plots all sources for which detections are available
for both axes.
The subset of sources that are
known binaries, with orbital separations $<2.5\arcsec$ (Table 1),
are shown as gray-brown filled points
(White \& Ghez 2001; Prato et al.\ 2002;
Mathieu, Mart\'in \& Magazzu 1996; and references therein).
The submillimeter observations do not resolve the binaries, so the 
reported fluxes may may contain contributions from circumstellar 
disks as well as a circumbinary disk. 
The presence of an orbiting stellar companion may have a significant
impact on the disc properties, e.g., by truncating a disc and thereby
both reducing the disc mass and altering the SED
(Jensen, Mathieu \& Fuller 1996).
Indeed, the median disc mass for the binary population is
$\log \Mdisk = -2.8$, lower than the median disc mass
of $\log \Mdisk = -2.3$ for the entire sample shown.
Since a binary companion
introduces additional complexity and may obscure important trends
that result from other physical processes,
we exclude these systems from our analysis.
The remaining filled black points indicate sources that have no known
binary companion within $2.5\arcsec$.

In the lower panel, the non-binary systems are further divided
into those that have a transition-like SED (crosses), as described
in \S 2.4, and those that do not (filled black points).
The sources with transition-like SEDs are:
GO Tau, LkCa 15, DN Tau, DM Tau, GM Aur,
V836 Tau, UX Tau A, and CX Tau
in order of decreasing disc mass.
The SEDs of GO Tau and DN Tau (Furlan et al.\ 2006)
are similar to those of the well studied transition objects 
GM Aur and DM Tau in that they also lack strong infrared excesses at 
short wavelengths but have optically thick outer discs.  
This is also the case for
V836 Tau, which was originally classified as a transition object
by Strom et al.\ (1989; see also Padgett et al.\ 2006).
UX Tau A is noted as a transition object by Furlan et al.\ (2006).
LkCa 15 has an SED that indicates an inner disc ``gap'' of a few AU
(Bergin et al.\ 2004).  

The concentration of crosses toward the lower right region of the plot
suggests that transition objects inhabit a restricted region of the  
$\Mdotstar$ vs. $\Mdisk$ plane.  
The existence of identifiable demographic groups demonstrates 
that the random errors in stellar accretion rates and disc masses
(\S 3.2 and 3.2) are apparently insufficient to obscure significant
population trends.
We can estimate statistically the likelihood that the transition
objects represent a distinct population in terms of their disc masses 
and accretion rates.  Figure 4 compares the distributions of stellar
accretion rates (top) and disc masses (bottom) for transition objects
(dashed histograms) and non-transition single stars (solid
histograms).  For sparse data of this kind, the two-sided KS test
is a good way to estimate the likelihood that the two distributions
are drawn from the same parent distribution.
With this approach, we find that there is a less than 3\% probability that the
stellar accretion rates of the transition and non-transition objects
are drawn from the same parent distribution.  
For the disc masses, there is a less than 12\% probability 
that the disc masses of the transition and non-transition objects are
drawn from the same parent distribution.

A more useful estimate would jointly compare the distributions in
both $\Mdotstar$ and $\Mdisk$.
A two-dimensional two-sided KS test that accounts for the distribution
of the two samples in both quantities estimates that there is a 
less than 0.8\% probability that the transition and non-transition
objects are drawn from the same parent distribution.
Thus, even with the limited data available at present, it appears  
that the transition objects represent a distinct population in terms of
their stellar accretion rates and disc masses.

For the stellar accretion rates, the median of the distribution 
is $\log \Mdotstar = -8.5$ for the transition objects and
$\log \Mdotstar = -7.6$ for the non-transition objects.
For the disc masses, the median of the distribution
is $\log \Mdisk = -1.6$ for the transition objects and
$\log \Mdisk = -2.2$ for the non-transition objects.
Thus, compared to non-transition single stars, transition objects are
biased to lower stellar accretion rates and larger disc masses. 
We discuss the possible origin of these biases in \S 4.

The likelihood that transition objects represent a 
population distinct from other classical T Tauri stars 
has possible implications for our understanding  
of the large dispersion in the $\Mdotstar$ vs.\ $\Mdisk$ plane. 
In simple disc models the disc accretion rate
$\Mdotd \propto \nu\Sigma$ locally, where
$\Sigma$ is the disc mass surface density and
$\nu =\alpha c_s H$ is the
viscosity parametrized in terms of the sound speed $c_s$ and the
disc scale height $H$.
If all discs have a similar sizes (e.g., outer disc radii) and 
radial mass distributions
(e.g., in a steady accretion disc $\Sigma \propto r^{-1}$;
Hartmann et al.\ 1998),
$\Sigma$ would be proportional to $\Mdisk$.
If, in addition, the disc accretion rate equals the stellar
accretion rate, we would expect $\Mdotstar \propto \Mdisk$.
In contrast, Figure 3 (top) shows considerable scatter with no
apparent trend between the two quantities.
The apparent lack of the expected correlation has been noted
previously (e.g., Dullemond, Natta, \& Testi 2006).

If the transition objects represent a distinct population
whose stellar accretion rates have been modified downward from
their initial values (see \S 4), this downward evolution can account
for some of the dispersion in the
$\Mdotstar$ vs.\ $\Mdisk$ plane.
To illustrate this, we show least absolute deviation fits to the
distributions in Figure 3.
In comparison to a minimized chi-square error criterion,
least absolute deviation fits are less sensitive to statistical
outliers.
In the top panel of Figure 3, a least absolute deviation fit to the
$\Mdotstar$ vs.\ $\Mdisk$ values for all sources for which there are
detections in both axes gives a slope of 0.22, relatively far from the linear
relation that would be expected in simple disc models.

If we remove the binaries from the sample (bottom panel) and fit
separately the slopes of the the non-transition single stars
(solid line) and the transition objects (dashed line),
both populations
have a slope of 0.94, close to the value of unity that would
be expected if mass accretion rate is proportional to disc mass.
The dashed line is offset downward from the solid line by approximately
an order of magnitude, similar to the difference in the median
accretion rates between the transition objects and the non-transition
single stars.
Because of the small number of sources that make up the transition
object sample, the slope for that population is not well-determined.

\setcounter{figure}{2}
\begin{figure}
\includegraphics[scale=0.5]{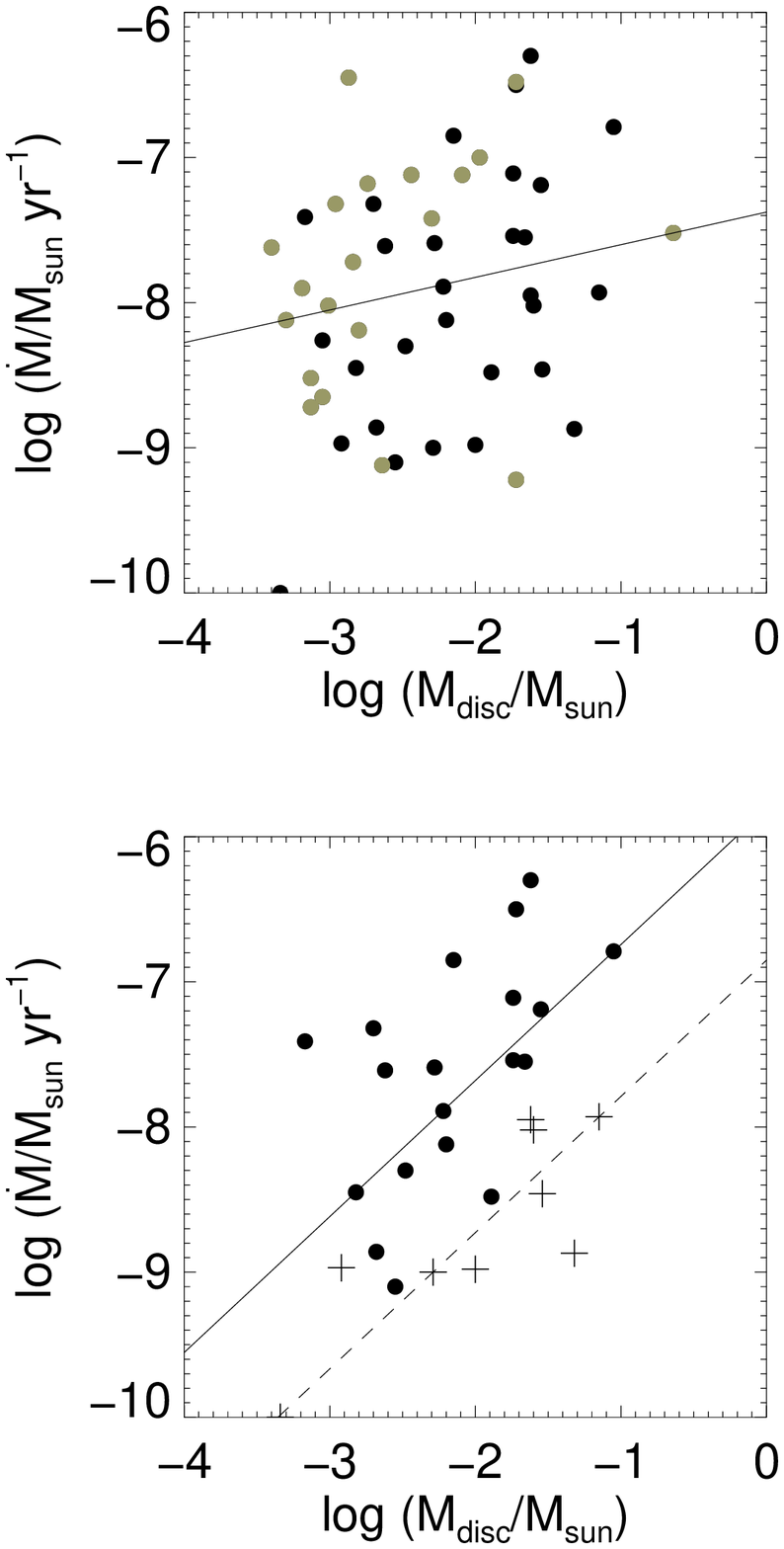}
\caption{Stellar accretion rates and outer disc masses for
sources in Taurus-Auriga.  Top panel: systems possessing known a
binary companion within $2.5\arcsec$ (gray-brown points) have lower
median disc masses compared to systems without such binary
companions (black points).  A least absolute deviation fit to
all of the points shown gives a slope of 0.22 (solid line).
Bottom panel: among the non-binary systems, those with transition-like
SEDs (crosses) are segregated from those with non-transition
SEDs (black points), with the transition objects having lower stellar
accretion rates for their disc masses.
Separately fitting the single, non-transition stars (solid line)
and the transition objects (dashed line) gives a slope of 0.94
for each population.
The upper limit to the accretion rate determined for
CoKu Tau/4 is included in the fit for the transition objects. 
}
\end{figure}

\section{Discussion} 

We have found that transition objects in Taurus occupy a restricted region
of the $\Mdotstar$ vs.\ $\Mdisk$ plane.  Compared to non-transition
single stars in Taurus, they have stellar accretion rates that are
typically $\sim 10$ times lower at the same disc mass.
In addition, the median disc mass for the transition objects is
$\sim 4$ times larger than the median disc mass for the
non-transition single stars.  The results place useful constraints
on the possible explanations for the origin of the transitional SEDs
in these systems.  They also shed some light on the nature of the 
relation between disc masses and disc accretion rates.
For example, the observed properties allow us to sort among the
following proposed explanations for transitional SEDs.  

(1) {\it Grain growth and the formation of rocky planetary cores in the
inner disc} (e.g., Strom et al.\ 1989; Dullemond \& Dominik 2005).
If small grains are consumed in this process, the inner disc will
become optically thin while remaining gas rich.  As a result, the
accretion rate through the disc and on to the star will continue
unabated.  

(2) {\it The dynamical clearing of a large gap by a Jovian mass planet}
(e.g., Skrutskie et al\ 1990; Marsh \& Mahoney 1992; 
Calvet et al.\ 2002; Rice et al.\ 2003; D'Alessio et al.\ 2005;
Calvet et al.\ 2005; Quillen et al.\ 2004).  
When a giant planet forms in a disc with a mass sufficient
to open a gap, gap crossing streams, from the outer disc 
(beyond $\Rhole$) to the planet, and from the planet to the 
inner disc (within $\Rinner$), are expected to allow 
continued accretion on to both planet and star, the latter through
the replenishment of an inner disc (Lubow, Seibert \& Artymowicz 1999; 
see also Kley 1999; 
Bryden et al.\ 1999; 
D'Angelo, Henning, \& Kley 2002;  
Bate et al.\ 2003; Lubow \& D'Angelo 2006).  
By the time the planet is massive enough to open a gap, it is also
able to divert a significant fraction of the mass accreting from
the outer disc on to itself.  Thus the stellar accretion rate is 
predicted to be reduced by a factor $\sim 0.1$ 
(Lubow \& D'Angelo 2006)
to $\sim 0.05$ (Varni\`ere et al.\ 2006)
compared to the mass accretion rate through the outer disc.

(3) {\it The isolation of the outer disc by a supra-Jovian mass planet,
followed by the viscous draining of the inner disc.}  As the
planet grows in mass via accretion from the gap-crossing streams,
accretion past the planet effectively ceases
(Lubow et al.\ 1999), isolating the inner and outer disc.
When the remaining inner disc material accretes on to the star,
the system is left with a large inner hole,
devoid of both gas and dust, and no
further stellar accretion.

(4) {\it Disk photoevaporation.}  Irradiation by energetic
(UV and X-ray) photons from the star heat
the disc surface layers to high temperatures $\sim 10^4$\,K.
Beyond a radius of $\sim$7--10\,AU where the thermal velocity of the
surface layers exceeds the escape speed ($\sim 10\,\kms$),
a photoevaporative flow (characterized by a mass loss rate of
$\sim 4\times 10^{-10}\,\Msunperyr$; Hollenbach, Yorke \& Johnstone 2000)
is expected to develop.
As discs viscously spread and accrete, eventually the column density
of the outer disc drops sufficiently that the disc accretion rate
becomes comparable to the disc photoevaporation rate.
At this point, the region beyond the photoevaporation radius
is losing material via photoevaporation more rapidly
than it can resupply the inner disc through viscous
accretion.  The inner disc is thereby decoupled from the outer
disc.  Material in the inner disc accretes on to the star,
leaving behind a large inner hole,
devoid of gas and dust (the ``UV Switch Model''; Clarke et al.\
2001), producing a transitional SED.

\setcounter{figure}{3}
\begin{figure}
\includegraphics[width=65mm]{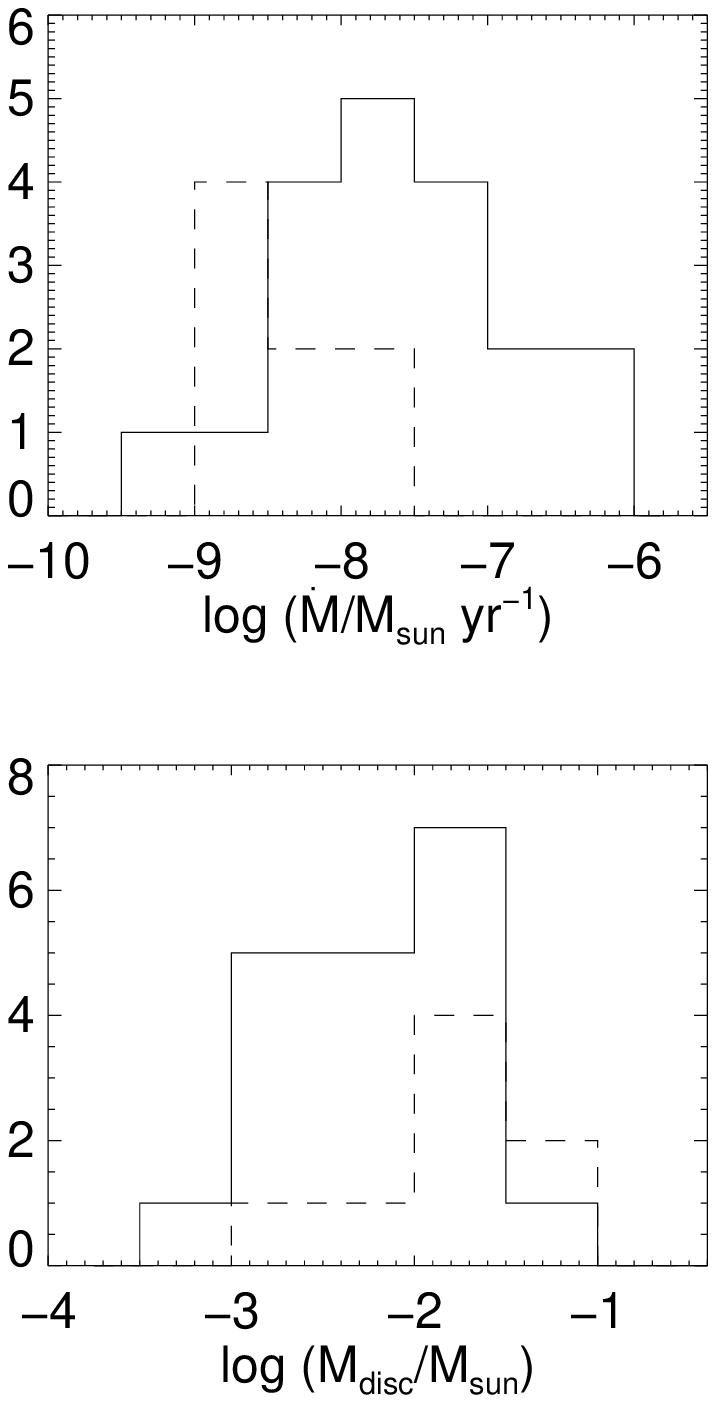}
\caption{Comparison of the distributions of stellar accretion rates 
(top) and disc masses (bottom) for transition objects (dashed line) 
and single, non-transition stars (solid line) in the Taurus star
forming region.  Compared to single, non-transition stars,
transition objects are biased to lower stellar accretion rates
and larger disc masses.
}
\end{figure}

When compared with these scenarios, the factor of $\sim 10$ decrement
in the observed stellar accretion rates of transition objects
compared to those of non-transition single stars at a given disc mass
is remarkably similar to predictions for the Jovian mass planet formation 
scenario described above.
The higher average disc masses for the transition objects,
compared to those of non-transition single stars may also be consistent
with this scenario since
more massive discs are expected to be conducive to the
formation of giant planets, either in the core accretion scenario
(e.g., Lissauer \& Stevenson 2006) or
if gravitational instabilities play a role (e.g., Boss 2005;
Durisen et al.\ 2005).
These similarities are suggestive that the formation of Jovian mass
planets plays a role in explaining the origin of at least 
some transitional SEDs in Taurus.   

Indeed, for the Taurus transition objects with higher disc masses
($\ga 0.01\Msun$),  
it is relatively straightforward to rule out 
the UV Switch model as an explanation for the transitional SEDs.
The high disc masses of these systems imply large 
gas column densities at the gravitational radius, suggesting that
photoevaporation can hardly be on the verge of decoupling the
inner and outer discs. 
The stellar accretion rates of these sources
($10^{-9} - 10^{-8}\Msunperyr$) are in excess of
the nominal disc photoevaporation rate of $4\times 10^{-10}\Msunperyr$,
further arguing against photoevaporation as the origin of the
transitional SED. 
Since disk photoevaporation rates are poorly known, sources 
with stellar accretion rates somewhat larger than the nominal 
$4\times 10^{-10}\Msunperyr$
could be on the verge of clearing their inner disks.  However, 
the inner disk is expected to clear rapidly once the 
stellar accretion rate drops down to the disk photoevaporation 
rate (Clarke et al.\ 2001); as a result, we would expect that 
few of the sources with stellar accretion rates of 
$\sim 10^{-9} \Msunperyr$ are passing through this 
brief evolutionary phase.  The high disk masses of these 
transition objects ($\ga 0.01\Msun$) support this interpretation.
Similarly, large accretion rates ($10^{-9} - 10^{-8}\Msunperyr$) 
are not expected 
if these systems have formed a supra-Jovian mass planet.
In contrast, significant accretion rates {\it are} expected when discs form
planetesimals.  
However, unlike the presence of a Jovian mass planet, this mechanism
does not provide a simple explanation for the low accretion
rates of these objects for their disc masses. 

The presence of an orbiting Jovian mass planet has indeed been the
favored explanation for the SEDs of the Taurus transition objects
GM Aur and DM Tau (Marsh \& Mahoney 1992; Rice et al.\ 2003;
Bergin et al.\ 2004; Calvet et al.\ 2005).
If such a planet were present, with a mass large enough to reduce
the accretion rate on to the star,
we would infer artificially small values of $\alpha$  
for the outer disc if we assumed that the accretion rate through the 
outer disc equalled the observed stellar accretion rate.
Indeed, a low value of the viscosity parameter $\alpha \sim 0.001$
is inferred for DM Tau under this assumption (Calvet et al.\ 2005).
Instead, if a Jovian mass planet is present,
most of the mass accreting through the outer disc may accrete
on to the planet rather than the star.
If the outer disc accretion
rate is then $\sim 10$ times the stellar accretion rate
(Lubow \& D'Angelo 2006), the inferred value of $\alpha$ characterizing
the outer disc of DM Tau would be $\sim 10$ times larger,
i.e., more similar to the value of $\alpha \simeq 0.01$ that is
believed to be typical of T Tauri discs (Hartmann et al.\ 1998).

Notably, at least one of the transition objects in Taurus
(CoKu Tau/4) has a disc mass
($\la 10^{-3}\,\Msun$) and
a stellar accretion rate
($\la 4\times 10^{-10}\Msunperyr$) that are both low enough
that its transitional SEDs may result from disc photoevaporation
rather than giant planet formation. 
CX Tau is a possibly similar case, although its estimated disk mass 
and stellar accretion rate are somewhat larger than the above limits.
While photoevaporation is not expected to erode away the inner regions of 
high mass discs ($\ga 0.01\Msun$) until they are 6--15 Myr old
(Clarke et al.\ 2001; Alexander et al.\ 2006), discs with sufficiently low
initial masses can photoevaporate away on much shorter timescales, 
comparable to the age of Taurus. 
For example, scaling the standard accretion disc model of
Hartmann et al.\ (1998) down in mass
shows that a disc with $\alpha=10^{-2}$ and $M_d=10^{-3}\Msun$
would spread beyond 100\,AU and evolve to
$\Mdotstar < 3\times 10^{-10}\Msunperyr$
on a time scale of $\sim 1$ Myr, a mass accretion rate low enough
that photoevaporation would be able to create an inner hole.

If we interpret CX Tau and CoKu Tau/4 as systems whose SEDs are
the result of photoevaporation, the remaining number of transition
objects is roughly consistent with the interpretation that
they are systems undergoing giant planet formation.
The fraction of such transition objects is 7 out of a total of
28 non-binary T Tauri stars or 25\%.  This is larger than the
5-10\% of nearby FGK stars currently known to harbor extrasolar
giant planets.
If the current detection statistics are extrapolated to 20 AU,
the expected fraction of planet-bearing systems rises to $\sim 12$\%
(Marcy et al.\ 2005).  
This is not totally inconsistent with the 25\% fraction of
transition objects identified here, given both the small sample of 
objects studied and the possible role of 
planetary orbital migration in accounting
for any difference in the incidence rates of accreting transition
objects and extrasolar giant planets.

Perhaps surprisingly, there is a dearth of transition objects
intermixed with the population of non-transition, single stars.
Systems that are forming planetesimals and protoplanets (i.e., 
objects too low in mass to open a gap) would be expected to
occupy this region of the 
diagram.  
That is, they would appear as systems with optically thin inner
discs but otherwise unaltered stellar accretion rates.
Perhaps the lack of such systems indicates that
planetesimal and protoplanet formation is inefficient in
clearing the dust from the inner regions of accreting discs
(cf.\ Dullemond \& Dominik 2005).
This might be the case if the accretion of small dust grains from 
larger disk radii is effective in ``filling in'' the opacity 
holes created by grain growth at smaller radii. 
Alternatively, the phase of planetesimal and protoplanet formation
may be very short-lived. 

A tempting counter-hypothesis is that systems that have experienced
significant grain growth and settling (and are forming planetesimals
and protoplanets) are intermixed with the transition objects we have
identified as candidate giant-planet forming systems.  This might
explain the diversity of transition object SEDs (Figure 2) that 
range from those with large ``dips'' (e.g., GM Aur) to 
``anemic'' SEDs (e.g., GO Tau) that might be more characteristic 
of systems that have undergone grain settling and growth.  
However, as already noted, it is difficult to understand why systems
that have experienced significant grain growth 
would show lower accretion rates for their disk masses.  By reducing
the small grain population in the surface layers of the disk, 
grain growth is expected to lead to larger ionization
fractions and therefore larger surface disk columns that are
susceptible to the magnetorotational instability (e.g., Sano et
al.\ 2000).  We would then naively expect that such disks would
have {\it larger} accretion rates for their disk masses, rather
than the reduced accretion rates that are observed.  Whether or 
not this is in fact the case, 
we may be able to determine observationally whether transition 
objects produced via grain growth are intermixed with systems that 
have formed giant planets,  
by measuring the radial distribution of gas in the disk. 
The radial distribution of gas would be continuous for the grain 
growth scenario, whereas it would show a radial gap if a 
Jovian mass planet has formed.   

Finally, we note that if we interpret the transition objects
as systems in which the stellar accretion rate has been reduced
from its original value (either by the presence of a giant 
planet or by photoevaporation) so that the accretion rate on to 
the star no longer traces the outer disc accretion rate, 
this implies a reduced scatter in the relation between
disc mass and accretion rate. 
If we consider only the single, non-transition stars as the
systems in which  
the stellar accretion rate might more closely trace the
{\it disc\/} accretion rate (Figure 3, bottom),
the range in $\Mdotstar$ at a given $\Mdisk$ is reduced by a factor
of $\sim 3$ compared to the range when the transition objects are
included.

By considering only single, non-transition objects, 
the relation between disc mass and stellar accretion rate has 
steeper a slope compared to that in Figure 3 (top), quite close to
the expected value of unity.   
There are many potential explanations for the remaining scatter.
Disks may have substantial dead zones that do 
not participate in accretion (Gammie 1996).   
Disk grain opacities may evolve (D'Alessio et al.\ 2006;
Natta et al.\ 2006), or the dust content 
that is measured by the submillimeter continuum may not reliably 
trace the gas (Takeuchi \& Lin 2005).  
Alternatively, 
the viscosity parameter $\alpha$ may not be a constant from
star to star, or  
discs may possess a range of initial angular momenta 
(Hartmann et al.\ 1998; Dullemond et al.\ 2006).

One potential problem with the interpretation that the transition
objects with higher disc masses have formed Jovian mass companions
is that their SEDs do not appear to show the distinctive
``gap'' structure one expects if a Jovian mass planet is present.
Several of these sources (GM Aur, DM Tau) have been fit with little
to no dust in the inner disc (e.g., Calvet et al.\ 2005), in marked
contrast to the highly optically thick inner disc (within
$\Rinner < \Rhole$) that would be expected for transition objects
with T Tauri-like accretion rates of $10^{-9}-10^{-8}\Msunperyr$.
This conundrum suggests the importance of developing a deeper
understanding of the processes that affect the vertical distribution 
of grains in the outer disk, the transport of grains from the outer 
to the inner disc, as well as the transformation of grain properties
in the inner disc (e.g., grain growth, destruction in spiral shocks)
in order to use SEDs as a finely honed tool for distinguishing among
the possibly physical origins of a transition SED.

Rice et al.\ (2006) have suggested that the filtering of
grains at gap edges can help to explain the lack of dust opacity in
the inner disc (within $\Rinner$).  As they describe, gas-grain 
coupling may drive grains
larger than 1--10$\micron$ away from the gap edge allowing only a
small fraction of
the solid mass to accrete past $\Rhole$.  Such a mechanism will
reduce the amount of solids reaching both the inner disc and
the planet in this stage of evolution.
As an additional mechanism, the high grain
densities that arise as material from the outer disc is
concentrated into narrow accretion streams may also give rise
to rapid grain growth and a further reduction in the disc opacity.
Similarly, the possible presence of a dead zone (Gammie 1996)
in the outer disc, combined with grain growth and settling out of
the surface regions of the disk (i.e., the region undergoing accretion), 
may also reduce the mass of small grains reaching $\Rhole$.
A dead zone has been invoked, similarly, by Ciesla (2006) in the 
context of {\it non}-accreting disks, where a transition-like SED 
may be produced in disks that have experienced significant grain 
growth but have not formed giant planets.  

An additional potential problem with the interpretation that the 
transition objects with higher disc masses have formed Jovian mass 
companions is that the mass doubling time for the putative 
accreting planet is quite short.  The median stellar accretion 
rate for these systems is $\sim 3 \times 10^{-9}\Msunperyr$, which
for a planetary accretion rate 10 times larger implies a mass 
doubling time of 0.03 Myr, an interval short compared
to the $\sim 1$ Myr age of Taurus.  
Such a mass accretion timescale might be consistent with the 
observed fraction of transition objects if sources in Taurus are 
approximately coeval, giant planet formation is common, and it takes 
close to 1 Myr to form a planet massive enough to open a gap.
Alternatively, 
other processes for forming transition disks may be at work. 
If so, they must provide a natural explanation for the 
apparent tendency for transition objects to have large disk 
masses and low stellar accretion rates.   

Given these uncertainties, it would be useful to identify
additional ways of confirming the presence of a giant planet beyond
the modeling of SEDs.
Perhaps one of the most convincing approaches would be to
search for direct evidence of the presence of such a planet.
If the stellar accretion rate in these systems is $\sim 0.1$
of the outer disc accretion rate, the planet then accretes
at a high rate, $\sim 9$ times the stellar accretion rate,
making the planet easier to detect.
Among the transition objects in Taurus,
GM Aur is the most promising system for such a search because
of the potentially large separation between the planet and star
($\sim$ 20\,AU or $0.13\arcsec$).  It also has
a large accretion rate. 
The stellar accretion rate of $10^{-8}\Msunperyr$ implies an
accretion rate of $\sim 9 \times 10^{-8}\Msunperyr$ on to the planet.
Extrapolating from the calculations of 
Hubickyj, Bodenheimer \& Lissauer (2005) and 
Papaloizou \& Nelson (2005), the accretion luminosity 
of the planet is expected to be substantial, $\sim 0.02\Lsun$.
The resulting modest contrast ratio between the accreting planet 
and the central star may greatly facilitate the detection of such 
an object.

\section{Summary}

Using the approach introduced here, a study of the demographics of 
accreting T Tauri stars, we find that 
transition objects inhabit restricted regions of
the $\Mdotstar$ vs. $\Mdisk$ plane.  
Compared to non-transition 
single stars in Taurus, they have stellar accretion rates that are 
typically $\sim 10$ times lower at the same disc mass. 
In addition, the median disc mass for the transition objects is
$\sim 4$ times larger than the median disc mass for the 
non-transition single stars.  
The decrement in the stellar accretion rates and the higher
average disc masses are suggestive that the formation of Jovian
mass planets plays a role in explaining the origin of 
some transitional SEDs in Taurus. 

For the transition objects with higher disc masses ($\ga 0.01\Msun$),
it seems plausible that a Jovian mass planet has created a large gap
and suppressed the accretion rate on to the star 
(Lubow \& D'Angelo 2006; Varni\`ere et al.\ 2006).   
The high accretion rate $\sim 9\Mdotstar$ inferred for the
Jovian mass planets in these systems would make an 
accreting giant planetary companion easier to detect using
either direct imaging or through infrared interferometry.
The remaining transition objects
have the low disc masses $\la 0.001\Msun$ expected 
for discs that have cleared inner holes through photoevaporation.
In contrast, the paucity of sources
with stellar accretion rates and disc masses similar 
to those of non-transition, single T Tauri stars 
leads us to question the formation of planetesimals or low mass
protoplanets as a likely pathway to a transitional SED. 
Planetesimal or low mass protoplanet formation may be unable to
completely remove small dust grains from the disc;
alternatively, this phase of evolution may be very short-lived.

Accreting transition objects that arise as a consequence 
of the formation of either planetesimals or a Jovian mass planet
present a significant theoretical challenge: how to sustain an
optically thin inner region while accreting material from a
dusty outer disc.  Understanding
the physical mechanisms that can sustain an optically thin
inner region in an accreting transition disc
represents a critical step toward understanding
the evolutionary state of these objects.

Finally, considering transition objects as a distinct
demographic group among accreting T Tauri stars, whose
stellar accretion rates have been altered as a
consequence of either giant planet formation or disc
photoevaporation, leads to a tighter relation between
disc masses and stellar accretion rates, with a slope
between the two quantities that is close to the value of
unity expected in theories of disc accretion.
These results, while suggestive, are based on a small sample
of objects.  Future observations aimed at determining disc masses,
stellar accretion rates, and SEDs for large samples of T Tauri stars
will be critical in improving our understanding
of the demographics and nature of transition objects.

\section*{Acknowledgments}
We are grateful to Lee Hartmann, Nuria  
Calvet, Doug Lin, and Geoff Bryden for stimulating and
insightful discussions on this topic.  
We also thank the anonymous referee for thoughtful comments 
that improved the manuscript.

\begin{table*}
 \centering
 \begin{minipage}{140mm}
  \caption{Taurus-Auriga Stellar Accretion Rates and Disk Masses.}
  \begin{tabular}{@{}llllrlr}
  \hline
Name & 
Spectral Type & 
Binarity$^{\rm a}$ &
SED$^{\rm b}$ &
$\log \Mdotstar^{\rm c}$ &
$\Mdotstar$ Ref. &
$\log M_d$$^{\rm d}$ \\
 \hline
AA Tau   & K7 &  S  &  C  &  -8.48 &  G98 & -1.89\\
Anon 1   & M0 &  S  &  W  &  $<$-8.56 & WG01 & $<$-3.40\\
BP Tau   & K7 &  S  &  C?  &  -7.54 &  G98 & -1.74\\
CI Tau   & K7 &  S  &  C  &  -7.19 & H98 & -1.55\\
CoKu Tau$/$4 & M1.5 & X &  T  &  $<$-10 & $^{\rm e}$ & -3.34\\
CW Tau   & K3 &  S  &  C  &  -7.61 & WG01 & -2.62\\
CX Tau   & M2.5 &  S  &  T  &  -8.97 & H98 &  -2.92\\
CY Tau   & M1 &  S  &  C  &  -8.12 & G98 &  -2.20\\
DD Tau A & M1 &  B &  C  &  -8.72 & WG01 & -3.13\\
DE Tau   & M2 &  S  &  C  &  -7.59 & G98 &  -2.28\\
DF Tau A & M0.5 &  B  &  C  &  -7.62 & WG01 & -3.40\\
DG Tau   & K7-M0 &  S  &  F  &  -6.30 & G00 & -1.62\\
DH Tau   & M1 &  S  &  C  &  -8.30 & H98 & -2.48\\
DK Tau   & K7 &  B  &  C  &  -7.42 & G98 & -2.30\\
DL Tau   & K7 &  S  &  C  &  -6.79 & WG01 & -1.05\\
DM Tau   & M1 &  S  &  T  &  -7.95 & H98 & -1.62\\
DN Tau   & M0 &  S  &  C/T  &  -8.46 & G98 & -1.54\\
DO Tau   & M0 &  S  &  C/F  &  -6.85 & G98 & -2.15\\
DP Tau   & M0 &  S  &  C/F  &  -7.88 & H98 & $<$-3.30\\
DQ Tau   & M0 &  B  &  C  &  -9.22 & G98 & -1.72\\
DR Tau   & K7 &  S  &  F  &  -6.50 & G00 & -1.72\\
DS Tau   & K5 &  S  &  C  &  -7.89 & G98 & -2.22\\
FM Tau   & M0 &  S  &  C  &  -8.45 & H98 & -2.82\\
FO Tau A & M2 &  B  &  T?  &  -7.90 & WG01 & -3.19\\
FQ Tau   & M2 &  B &  T? &  -6.45 & H98  & -2.87\\
FS Tau   & M1 &  B &  C/F &  -9.12 & WG01 & -2.64\\
FV Tau A & K5 &  B &  C  &  -7.32 & WG01 & -2.96\\
FX Tau   & M1 &  B &  C  &  -8.65 & H98 & -3.05\\
FY Tau   & K7 &  S  &  X  &  -7.41 & H98 & -3.17\\
FZ Tau   & M0 &  S  &  C  &  -7.32 & WG01 & -2.70\\
GG Tau Aa & K7 &  B  &  C  &  -7.52 & WG01 & -0.64\\
GH Tau A & M2 &  B &  C  &  -8.52 & WG01 & -3.13\\
GK Tau   & K7 &  B &  C  &  -8.19 & G98 & -2.80\\
GM Aur   & K3 &  S  &  T  &  -8.02 & G98 & -1.60\\
GO Tau   & M0 &  S  &  C/T  &  -7.93 & H98 & -1.15\\
Haro 6$-$37 & K6 & B  &  C  &  -7.00 & H98 & -1.97\\
HBC 376  & K7 &  S  &  X  &  $<$-8.92 & WG01 & $<$-3.54\\
HBC 388  & K1 &  S  &  W  &  $<$-8.03 & WG01 & $<$-3.49\\
HO Tau   & M0.5 &  S  &  C  &  -8.86 & H98 & -2.68\\
Hubble 4 & K7 &  S  &  W  &  $<$-7.78 & WG01 & $<$-3.36\\
IP Tau   & M0 &  S  &  C  &  -9.10 & G98 & -2.55\\
IQ Tau   & M0.5 &  S  &  C  &  -7.55 & H98 & -1.66\\
IS Tau A & K7 &  B &  C  &  -7.72 & WG01 & -2.84\\
L1551$-$51 & K7 &  S  &  W  &  $<$-9.12 & WG01 & $<$-3.19\\
L1551$-$55 & K7 &  S  &  X  &  $<$-9.32 & WG01 & $<$-3.56\\
LkCa 14  & M0 &  S  &  X  &  $<$-8.47 & WG01 & $<$-3.35\\
LkCa 15  & K5 &  S  &  C/T  &  -8.87 & H98 & -1.32\\
LkCa 19  & K0 &  S  &  X  &  $<$-9.62 & WG01 & $<$-3.30\\
LkCa 4   & K7 &  S  &  W  &  $<$-8.35 & WG01 & $<$-3.70\\
LkCa 5   & M2 &  S  &  W  &  $<$-9.62 & WG01 & $<$-3.72\\
RW Aur A & K3 &  B  &  C  &  -7.12 & WG01 & -2.44\\
RY Tau   & G1 &  S  &  C  &  -7.11 & C04 & -1.74\\
SU Aur   & G1 & X  &  C  &  -8.26 &  C04 & -3.05\\
T Tau A  & G6 & B &  F  &  -7.12 & WG01 & -2.09\\
UX Tau A & K2 &  S  &  T  &  -9.00 & $^{\rm f}$ & -2.29\\
UY Aur   & K7 &  B &  C  &  -7.18 & G98 & -2.74\\
UZ Tau E & M1 &  B &  C  &  -6.48 & VBJ93  & -1.72\\
V410 Tau A & K3 &  B  &  W  &  $<$-8.42 & WG01 & $<$-3.44\\
\hline
\end{tabular}
\end{minipage}
\end{table*}

\begin{table*}
 \centering
 \begin{minipage}{140mm}
  \contcaption{Taurus-Auriga Stellar Accretion Rates and Disk Masses.}
  \begin{tabular}{@{}llllrlr}
  \hline
Name & 
Spectral Type & 
Binarity$^{\rm a}$ &
SED$^{\rm b}$ &
$\log \Mdotstar^{\rm c}$ &
$\Mdotstar$ Ref. &
$\log M_d$$^{\rm d}$ \\
 \hline
V773 Tau & K3 &  B  &  C/T  &  $<$-9.62  & WG01 & -3.33\\
V807 Tau A & K7 &  B &  X  &  -8.02 & WG01 & -3.01\\
V819 Tau & K7 &  S  &  W$^{\rm g}$  &  $<$-8.48 & WG01 & $<$-3.35\\
V827 Tau & K7 &  S  &  W  &  $<$-8.15 & WG01 & $<$-3.50\\
V836 Tau & K7 &  S  &  T?  & -8.98 & HEG95 & -2.00\\
V955 Tau & K7 &  B &  C  &  -8.12 & WG01 & -3.30 \\
\hline
\end{tabular}
\medskip \\
$^{\rm a}$ Binarity where S$=$single; 
B$=$binary with orbital separation $< 2.5\arcsec$; \\
$^{\rm b}$ SED class where C$=$classical T Tauri star;
F$=$extremely active accretor or flat spectrum source;
T$=$transition object, i.e., no IR excess shortward of
$10\micron$ and a significant excess at longer wavelengths;
W$=$optically thin or no IR excess shortward of $30\micron$;
X$=$unknown.
Entries with a C$?$, T$?$ or c$/$T indicate uncertainty in the
true level of the short-wavelength IR excess either because of
variability (e.g., BP Tau, V836 Tau) or companions (e.g., V773 Tau). 
These designations are also used in cases of ambiguity between
a transition phenomenon and naturally lower near-IR excess emission around
later-type stars where the irradiation flux impinging on the disc is smaller
(e.g., FO Tau, FQ Tau, GO Tau). \\
$^{\rm c}$ Stellar accretion rate in units of $\Msunperyr$. \\
$^{\rm d}$ Disk masses in units of $\Msun$ from
Andrews \& Williams (2005). \\
$^{\rm e}$ $\Mdotstar$ upper limit estimated assuming no accretion
component is measurable at H$\alpha$. \\
$^{\rm f}$ $\Mdotstar$ estimated from an accretion model of
the H$\alpha$ profile shown in Alencar \& Basri (2000). \\
$^{\rm g}$ Furlan et al.\ (2006) note that a 2MASS companion
with unknown mid-infrared properties, which also lay in the long-low slit,
possibly accounts for the longer wavelength excess in the measured SED.
\end{minipage}
\end{table*}

\begin{table*}
 \centering
 \begin{minipage}{140mm}
  \caption{SED Decrements of Transition Objects.}
  \begin{tabular}{@{}lcll@{}}
  \hline
Name &
SED &
$\Delta_{3.5}$ &
$\Delta_{5.0}$ \\
    &
    &
dex &
dex \\
 \hline
CoKu Tau$/$4 & T & 0.4  & 0.7 \\
CX Tau   & T   &   0.4  & 0.3 \\
DM Tau   & T   &   0.2  & $>$0.3 \\
DN Tau   & C/T &   0.2  & 0.1 \\
GM Aur   & T &   0.5  & 0.4 \\
FO Tau A & T?  &   0.25  & 0.15 \\
FQ Tau   & T?  &   0.4  & 0.3 \\
GO Tau   & C/T &   0.3  & 0.25 \\
LkCa 15  & C/T &   0.2  & 0.2 \\
UX Tau A & T   &   0.15  & 0.25 \\
V773 Tau & C/T &   0.35  & 0.25 \\
V836 Tau & T?  &   0.35  & 0.05 \\
\hline
\end{tabular}
\end{minipage}
\end{table*}


\begin{thebibliography}{}

\bibitem[\protect\citeauthoryear{}{}]{} 
Alexander R. D., Clarke C. J., Pringle J. E. 2006, MNRAS, 369, 229

\bibitem[\protect\citeauthoryear{}{}]{} 
Andrews S. M., Williams J. P. 2005, ApJ, 631, 1134

\bibitem[\protect\citeauthoryear{}{}]{} 
Baraffe I., Chabrier G., Allard F., Hauschildt P. H. 1998, A\&A, 377, 
403

\bibitem[\protect\citeauthoryear{}{}]{} 
Bate M. R., Lubow S. H., Ogilvie G. I., Miller K. A.
2003, MNRAS, 341, 213

\bibitem[\protect\citeauthoryear{}{}]{} 
Bergin E., et al. 2004, ApJ, 614, L133

\bibitem[\protect\citeauthoryear{}{}]{} 
Boss A. P. 2005, ApJ, 629, 535

\bibitem[\protect\citeauthoryear{}{}]{} 
Bryden G., Chen X., Lin D. N. C., Nelson R. P., Papaloizou J. C. B.
1999, ApJ, 514, 344

\bibitem[\protect\citeauthoryear{}{}]{} 
Calvet N., D'Alessio P., Hartmann L., Wilner D., Walsh A., 
Sitko M. 2002, ApJ, 568, 1008 

\bibitem[\protect\citeauthoryear{}{}]{} 
Calvet N., Muzerolle J., Brice\~no C., Hern\'andez J., Hartmann L., 
Saucedo J. L., Gordon K. D. 2004, AJ, 128, 1294 (C04) 

\bibitem[\protect\citeauthoryear{}{}]{} 
Calvet N., et al.\ 2005, ApJ, 630, L185

\bibitem[\protect\citeauthoryear{}{}]{} 
Ciesla, F. J. 2006, ApJ, in press (astro-ph/0611811)

\bibitem[\protect\citeauthoryear{}{}]{} 
Clarke C. J., Gendrin A., Sotomayor M. 2001, MNRAS, 328, 485

\bibitem[\protect\citeauthoryear{}{}]{} 
D'Alessio P., Calvet N., Hartmann L., Lizano S., Cant\'o J. 
1999, ApJ, 527, 893

\bibitem[\protect\citeauthoryear{}{}]{} 
D'Alessio P. et al.\ 2005, ApJ, 621, 461

\bibitem[\protect\citeauthoryear{}{}]{} 
D'Alessio P., Calvet N., Hartmann L., Franco-Hern\'andez R.,
Serv\'in H. 2006, ApJ, 638, 314

\bibitem[\protect\citeauthoryear{}{}]{} 
D'Angelo G., Henning T., Kley W. 2003, ApJ, 599, 548

\bibitem[\protect\citeauthoryear{}{}]{} 
D'Antona F., Mazzitelli I. 1998, 
in R. Rebolo, E. L. Martin, M. R. Zapatero Osorio, eds,
ASP Conf. Ser. 134, 
Astron. Soc. Pac., San Francisco, p. 442

\bibitem[\protect\citeauthoryear{}{}]{} 
Dullemond C. P., Dominik C. 2005, A\&A, 434, 971

\bibitem[\protect\citeauthoryear{}{}]{} 
Dullemond C. P., Natta A., Testi L. 2006, ApJ, 645, L69

\bibitem[\protect\citeauthoryear{}{}]{} 
Durisen R. H., Cai K., Mej\'ia A. C., Pickett M. K. 2005,
Icarus, 173, 417 

\bibitem[\protect\citeauthoryear{}{}]{} 
Furlan E., et al.\ 2006, ApJS, 165, 568 

\bibitem[\protect\citeauthoryear{}{}]{} 
Gammie C. F. 1996, ApJ, 457, 355

\bibitem[\protect\citeauthoryear{}{}]{} 
Gauvin L. S., Strom K. M. 1992, ApJ, 385, 217

\bibitem[\protect\citeauthoryear{}{}]{} 
Gullbring E., Hartmann L., Brice\~no C., Calvet N.
1998, ApJ, 492, 323 (G98)

\bibitem[\protect\citeauthoryear{}{}]{} 
Gullbring E., Calvet N., Muzerolle J., Hartmann L. 2000, ApJ, 544, 927 (G00)

\bibitem[\protect\citeauthoryear{}{}]{} 
Hartigan P., Edwards S., Ghandour L. 1995, ApJ, 452, 736 (HEG95)

\bibitem[\protect\citeauthoryear{}{}]{} 
Hartmann L., Calvet N., Gullbring E., D'Alessio P. 1998,
ApJ, 495, 385 (H98)

\bibitem[\protect\citeauthoryear{}{}]{} 
Hartmann L., Megeath S. T., Allen L., Luhman K., Calvet N., 
D'Alessio P., Franco-Hernandez R., Fazio G. 2005, ApJ, 
629, 881

\bibitem[\protect\citeauthoryear{}{}]{} 
Hartmann L., D'Alessio P., Calvet N., Muzerolle J. 2006, ApJ, 648, 484

\bibitem[\protect\citeauthoryear{}{}]{} 
Hollenbach D. J., Yorke H. W., Johnstone D. 2000, in
Mannings, V., Boss, A. P., Russell, S. S., eds,  
Protostars and Planets IV, University of Arizona, Tucson, p. 401

\bibitem[\protect\citeauthoryear{}{}]{} 
Hubickyj O., Bodenheimer P., Lissauer J. J. 2005, Icarus,
179, 415

\bibitem[\protect\citeauthoryear{}{}]{} 
Jensen E. L. N., Mathieu R. D., Fuller G. A. 1996,
ApJ, 458, 312

\bibitem[\protect\citeauthoryear{}{}]{} 
Kenyon S. J., Hartmann L. 1995, ApJS, 101, 117

\bibitem[\protect\citeauthoryear{}{}]{} 
Kley W. 1999, MNRAS, 303, 696

\bibitem[\protect\citeauthoryear{}{}]{} 
Lissauer J. J., Stevenson D. J. 2006, in 
B. Reipurth, ed, Protostars and Planets V, University of Arizona Press, 
Tucson, in press

\bibitem[\protect\citeauthoryear{}{}]{} 
Lubow S. H., Seibert M., Artymowicz P. 1999, ApJ, 526, 1001

\bibitem[\protect\citeauthoryear{}{}]{} 
Lubow S. H., D'Angelo G. 2006, ApJ, 641, 526

\bibitem[\protect\citeauthoryear{}{}]{} 
Marcy G., Butler R. P., Fischer D., Vogt S., Wright J. T.,
Tinney C. G., Jones H. R. A. 2005, Prog. Theor. Phys. Supp.,
158, 2005 

\bibitem[\protect\citeauthoryear{}{}]{} 
Marsh K. A., Mahoney M. J. 1992, ApJ, 395, L115

\bibitem[\protect\citeauthoryear{}{}]{} 
Mathieu R. E., Mart\'in E. L., Magazzu A. 1996, BAAS, 188, 60.05

\bibitem[\protect\citeauthoryear{}{}]{} 
McCabe C., Ghez A. M., Prato L., Duch\^ene G., Fisher R. S.,
Telesco C. 2006, ApJ, 636, 932

\bibitem[\protect\citeauthoryear{}{}]{} 
Muzerolle J., Calvet N., Hartmann L. 2001, ApJ, 550, 944

\bibitem[\protect\citeauthoryear{}{}]{} 
Muzerolle J., Hillenbrand L., Calvet N., Brice\~no C., 
Hartmann L. 2003, ApJ, 592, 266

\bibitem[\protect\citeauthoryear{}{}]{} 
Muzerolle J., et al.\ 2006, ApJ, 643, 1003

\bibitem[\protect\citeauthoryear{}{}]{} 
Natta A., Testi L., Muzerolle J., Randich S., Comer\'on F.,
Persi P. 2004, A\&A, 424, 603

\bibitem[\protect\citeauthoryear{}{}]{} 
Natta A., Testi L., Calvet N., Henning T., Waters R.,
Wilner D. 2006, in 
B. Reipurth, ed, Protostars and Planets V, University of Arizona Press, 
Tucson, in press

\bibitem[\protect\citeauthoryear{}{}]{} 
Padgett D. L., et al.\ 2006, ApJ, 645, 1283

\bibitem[\protect\citeauthoryear{}{}]{} 
Palla F., Stahler S. 1999, ApJ, 525, 772 

\bibitem[\protect\citeauthoryear{}{}]{} 
Papaloizou J. C. B., Nelson R. P. 2005, A\&A, 433, 247

\bibitem[\protect\citeauthoryear{}{}]{} 
Prato L., Simon M., Mazeh T., McLean I. S., Norman D., 
Zucker S. 2002, ApJ, 569, 863

\bibitem[\protect\citeauthoryear{}{}]{} 
Quillen A. C., Blackman E. G., Frank A., Varni\`ere P.
2004, ApJ, 612, L137

\bibitem[\protect\citeauthoryear{}{}]{} 
Rice W. K. M., Wood K., Armitage P. J., Whitney B. A., 
Bjorkman J. E. 2003, MNRAS, 342, 79

\bibitem[\protect\citeauthoryear{}{}]{} 
Rice W. K. M., Armitage P. J., Wood K., Lodato G.
2006, astro-ph/0609808

\bibitem[\protect\citeauthoryear{}{}]{} 
Sano, T., Miyama, S. M., Umebayashi, T., Nakano, T. 
2000, ApJ, 543, 486 

\bibitem[\protect\citeauthoryear{}{}]{} 
Sicilia-Aguilar A., et al.\ 2006, 638, 897

\bibitem[\protect\citeauthoryear{}{}]{} 
Skrutskie M. F., Dutkevitch D., Strom S. E., Edwards S.,
Strom K. M., Shure M. A.  1990, AJ, 99, 1187

\bibitem[\protect\citeauthoryear{}{}]{} 
Strom K. M., Strom S. E., Edwards S., Cabrit S., 
Skrutskie M. F. 1989, AJ, 97, 1451

\bibitem[\protect\citeauthoryear{}{}]{} 
Takeuchi T., Lin D. N. C. 2005, ApJ, 623, 482

\bibitem[\protect\citeauthoryear{}{}]{} 
Varni\`ere P., Blackman E. G., Frank A., Quillen A. C. 2006,
ApJ, 640, 1110

\bibitem[\protect\citeauthoryear{}{}]{} 
White R. J., Ghez A. M. 2001, ApJ, 556, 265 (WG01)

\bibitem[\protect\citeauthoryear{}{}]{} 
White R. J., Hillenbrand L. A. 2004, ApJ, 616, 998


\bibitem[\protect\citeauthoryear{}{}]{} 
Wolk S. J., Walter F. M. 1996, AJ, 111, 2066

\end{thebibliography}
\end{document}